\documentclass[pre,amssymb,amsmath,preprintnumbers,nofootinbib, 11pt]{revtex4-1}

\usepackage{graphicx}
\usepackage{color}
\usepackage{verbatim}
\usepackage{subfigure}
\usepackage{amssymb}

\begin{document}

\title{Modeling Maxwell's demon with a microcanonical Szilard engine}

\author{Suriyanarayanan Vaikuntanathan$^1$ and Christopher Jarzynski$^{1,2}$}
\affiliation{$^1$Chemical Physics Program, Institute for Physical Science and Technology,University of Maryland, College Park, MD 20742\\ 
$^2$Department of Chemistry and Biochemistry, University of Maryland, College Park, MD 20742}

\begin{abstract}
Following recent work by Marathe and Parrondo [PRL, {\bf 104}, 245704 (2010)], we construct a classical Hamiltonian system whose energy is reduced during the adiabatic cycling of external parameters, when initial conditions are sampled microcanonically.
Combining our system with a device that measures its energy, we propose a cyclic procedure during which energy is extracted from a heat bath and converted to work, in apparent violation of the second law of thermodynamics.
This paradox is resolved by deriving an explicit relationship between the average work delivered during one cycle of operation, and the average information gained when measuring the system's energy.
\end{abstract}

\maketitle

\section{Introduction}
\label{sec:intro}

The Kelvin-Planck statement of the second law of thermodynamics asserts that
no process is possible whose sole result is the extraction of energy from a heat bath, and the conversion of that energy into work.~\cite{Finn1993}
Because this statement is formulated in terms of energy rather than entropy, it provides an attractive starting point for exploring the microscopic foundations of the second law.
This is particularly true when we consider an immediate corollary of the Kelvin-Planck statement: when a thermally isolated system, initially in equilibrium, evolves under a cyclic variation of external parameters, its internal energy cannot decrease.\footnote{
If its energy were to decrease, then at the end of the process the system could be returned to its initial state by equilibrating it with a heat bath at temperature $T$, resulting in the net conversion of heat to work.
}
Since an isolated system exchanges no heat with its surroundings, and is governed by familiar equations of motion -- Hamiltonian dynamics in the classical case, or the Schr\" odinger equation for a non-relativistic quantum system -- relatively few theoretical tools are needed to embark on an investigation of this statement.

Let us formulate the problem as follows.
A finite, classical system is described by a Hamiltonian $H({\bf z};\vec\lambda)$, where ${\bf z}=(q,p)$ denotes a point in $2D$-dimensional phase space, and $\vec\lambda = (\lambda_1,\cdots,\lambda_n)$ is a set of externally controlled parameters.
At time $t=0$ the system's initial conditions are sampled from an equilibrium distribution $p^{\rm eq}({\bf z})$, and then for $0\le t\le\tau$ the system evolves under Hamilton's equations as the parameters are made to trace out a closed loop in $\vec\lambda$-space.
We will use the notation $\vec\lambda_c(t)$ to denote such a cyclic protocol for varying the parameters, beginning and ending at $\vec\lambda^A \equiv \vec\lambda_c(0) = \vec\lambda_c(\tau)$.
The work performed on the system during this process is the net change in the value of the Hamiltonian,
\begin{equation}
W = H({\bf z}_\tau;\vec\lambda^A) - H({\bf z}_0;\vec\lambda^A),
\end{equation}
where the trajectory ${\bf z}_t$ describes the system's evolution from $t=0$ to $t=\tau$. 
Since Hamiltonian dynamics are deterministic, the value of $W$ is fully determined by the initial conditions: $W=W({\bf z}_0)$.
The Kelvin-Planck statement, viewed as a statistical prediction about averages, then implies the inequality,
\begin{equation}
\label{eq:kp}
\langle W \rangle \equiv \int {\rm d}{\bf z}_0 \, p^{\rm eq}({\bf z}_0) \, W({\bf z}_0) \ge 0 .
\end{equation}
We now ask, for what choices of the equilibrium distribution $p^{\rm eq}({\bf z})$ can this result be established rigorously?

When initial conditions are sampled from a canonical distribution
\begin{equation}
p_{can}^{\rm eq}({\bf z}) \propto \exp \left[ - \beta H_A({\bf z}) \right] \qquad , \qquad H_A({\bf z}) \equiv H({\bf z};\vec\lambda^A) ,
\end{equation}
Eq.~\ref{eq:kp} follows directly from the properties of Hamilton's equations~\cite{Jarzynski1997a,Allahverdyan2002,Campisi2008}.
In fact, this result extends to any distribution of initial conditions that is a decreasing function of energy~\cite{Allahverdyan2002,Campisi2008}.
Somewhat surprisingly, however, Eq.~\ref{eq:kp} is not universally valid when initial conditions are sampled from a microcanonical distribution,
\begin{equation}
p_{\mu can}^{\rm eq}({\bf z}) \propto \delta \left[ E_i-H_A({\bf z}) \right]
\end{equation}
This has been discussed by Allahverdyan and Nieuwenhuizen~\cite{Allahverdyan2002}, but to the best of our knowledge it was Sato~\cite{Sato2002} who first constructed a counter-example, involving a perturbed, one-dimensional harmonic oscillator.
For microcanonically sampled initial conditions, Sato described a cyclic variation of the Hamiltonian that results in a negative value of average work, $\langle W\rangle < 0$.
More recently, Marathe and Parrondo~\cite{Marathe2010} have developed another counterexample to Eq.~\ref{eq:kp}, involving a particle inside a box with hard walls and an insertable barrier.
For a given initial energy, Marathe and Parrondo describe a cyclic manipulation of the walls and the barrier, whose net effect is to reduce the energy of the system.
Ultimately, the particle can be brought arbitrarily close to zero kinetic energy by a succession of such cycles, with a different protocol for each cycle.

Inspired by Ref.~\cite{Marathe2010}, in the present paper we introduce and analyze another model system that violates Eq.~\ref{eq:kp}.
We consider a classical particle moving in a one-dimensional potential well, described by a pair of external parameters $\vec\lambda = (\lambda_L,\lambda_R)$ (see Eq.~\ref{eq:Hdef} and Fig.~\ref{fig:potential}).
We will discuss the design of protocols for varying these parameters cyclically with time, $\vec\lambda_c(t)$, in a manner that lowers the energy of the system.
In particular, for any choice of initial particle energy $E_i$, we will construct a protocol (which depends on the value of $E_i$) that reduces the particle's kinetic energy arbitrarily close to zero in a single cycle, bringing the system to a final state in which the particle sits nearly motionless at the bottom of the potential well.
In effect, the system is cooled near to ``absolute zero'' temperature.

Our model, like those of Refs.~\cite{Sato2002,Marathe2010}, suggests that a perpetual-motion device of the second kind could be constructed, operating by the following steps.
\begin{enumerate}
\item
The system is brought into contact and allowed to equilibrate with a thermal reservoir at temperature $T$.
The reservoir is then removed.
\item
The energy of the now-isolated system is measured.
\item
The system is subjected to a cyclic protocol that reduces its kinetic energy close to zero (as discussed above).
\end{enumerate}
By repeatedly performing this sequence of steps, we obtain a scenario in which energy is systematically extracted from the reservoir (step 1) and delivered as work to the agent that carries out the cyclic protocol (step 3).
This is reminiscent of Maxwell's demon~\cite{Maxwell1871,Leff2003,Maruyama2009}, only here the demon's role is to implement a cyclic protocol $\vec\lambda_c(t)$ based on the measured energy of the system, instead of opening or closing a trapdoor based on the observed motion of nearby particles.
The key to exorcising the demon -- that is, to reconciling this scenario with the second law of thermodynamics -- is to recognize that the repeated measurements of energy in step 2 result in the accumulation of information.
In order for the device to satisfy the ``sole result'' stipulation of the Kelvin-Planck statement (see above), this information must eventually be erased.
As famously discussed by Landauer~\cite{Landauer1961}, and by Bennett~\cite{Bennett1982} in the context of Szilard's engine~\cite{Szilard1929} -- another incarnation of Maxwell's demon -- the erasure of information carries an unavoidable thermodynamic cost of $k_BT\ln 2$ per bit.
We will show by explicit calculation that this cost ultimately wipes out any gains made by our device:
in the process of erasing the accumulated information, all of the work harvested by the device is returned as heat to the thermal reservoir.

In Sec.~\ref{sec:model} we introduce our model and discuss protocols $\vec\lambda_c(t)$ that reduce the energy of the system.
In Sec.~\ref{sec:erasure} we discuss the average amount of work that is extracted per cycle, when carrying out the three-step procedure discussed above;
this amount depends on the precision with which the initial energy is measured in step 2.
Using Landauer's principle for the work that must eventually be expended to erase the accumulated information ($k_BT\ln 2$ per bit), we will show that this is no less than the work extracted in step 3, regardless of the precision with which the initial energy is measured.
Thus in the final accounting, after all the bits of information are reset to zero, the device is unable to deliver work and the second law is rescued from the demon.

\section{Model and Protocols}
\label{sec:model}

\begin{figure}[tbp]         
\includegraphics[scale=0.4,angle=0]{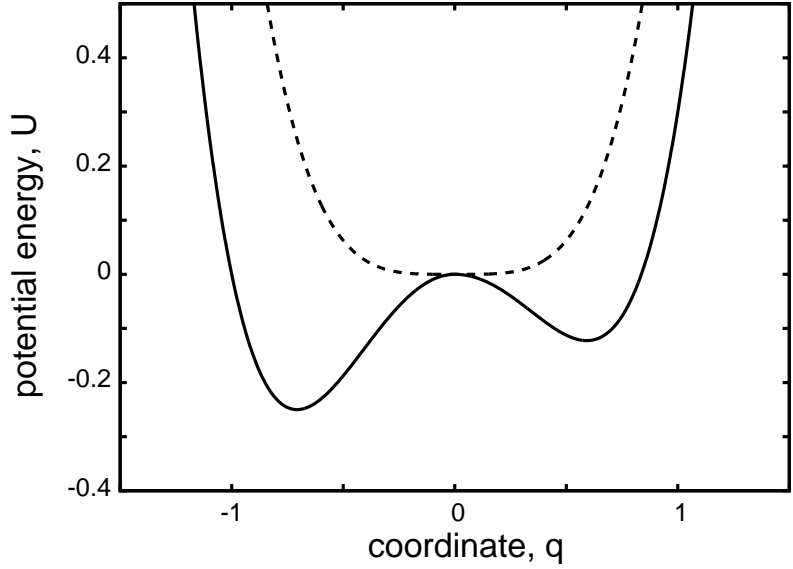} 
\caption{The solid curve depicts the potential $U(q;1.0,0.7)$, with local minima at $q_L^{\rm min}=-\sqrt{0.5}$ and $q_R^{\rm min}=+\sqrt{0.35}$ (see text).
The dashed curve is the unperturbed, quartic potential $U(q;0,0)$.
}
 \label{fig:potential}
\end{figure}

Consider a classical particle of unit mass moving in one dimension, governed by a Hamiltonian
\begin{equation}
\label{eq:Hdef}
H({\bf z};\vec\lambda) = \frac{p^2}{2} + U(q;\vec\lambda) \equiv
\frac{p^2}{2} + q^4 -
\begin{cases}
\lambda_L \, q^2 \quad & \text {if $q \le 0$} \\
\lambda_R \, q^2 \quad & \text {if $q \ge 0$}
\end{cases}
\end{equation}
where ${\bf z}=(q,p)$ is a point in the phase space of the particle, and $\vec\lambda=(\lambda_L,\lambda_R)$ is a point in two-dimensional parameter space, with $\lambda_L, \lambda_R \ge 0$.
The parameter $\lambda_L$ modulates the shape of the potential energy function in the region $q<0$:
when $\lambda_L > 0$, there is a local minimum at $q_L^{\rm min} = -\sqrt{\lambda_L/2}$, as illustrated in Fig.~\ref{fig:potential}.
Similarly, the value of $\lambda_R$ specifies a minimum at $q_R^{\rm min}=+\sqrt{\lambda_R/2}$.
We will refer to these regions as the left well and the right well.
When $\vec\lambda = (0,0)$, the particle moves in a quartic potential, which we call the unperturbed system.

\begin{figure}[tbp]
   \subfigure[\, Symmetric protocol.]{
   \label{fig:symmetricProtocol}
   \includegraphics[scale=0.25,angle=0]{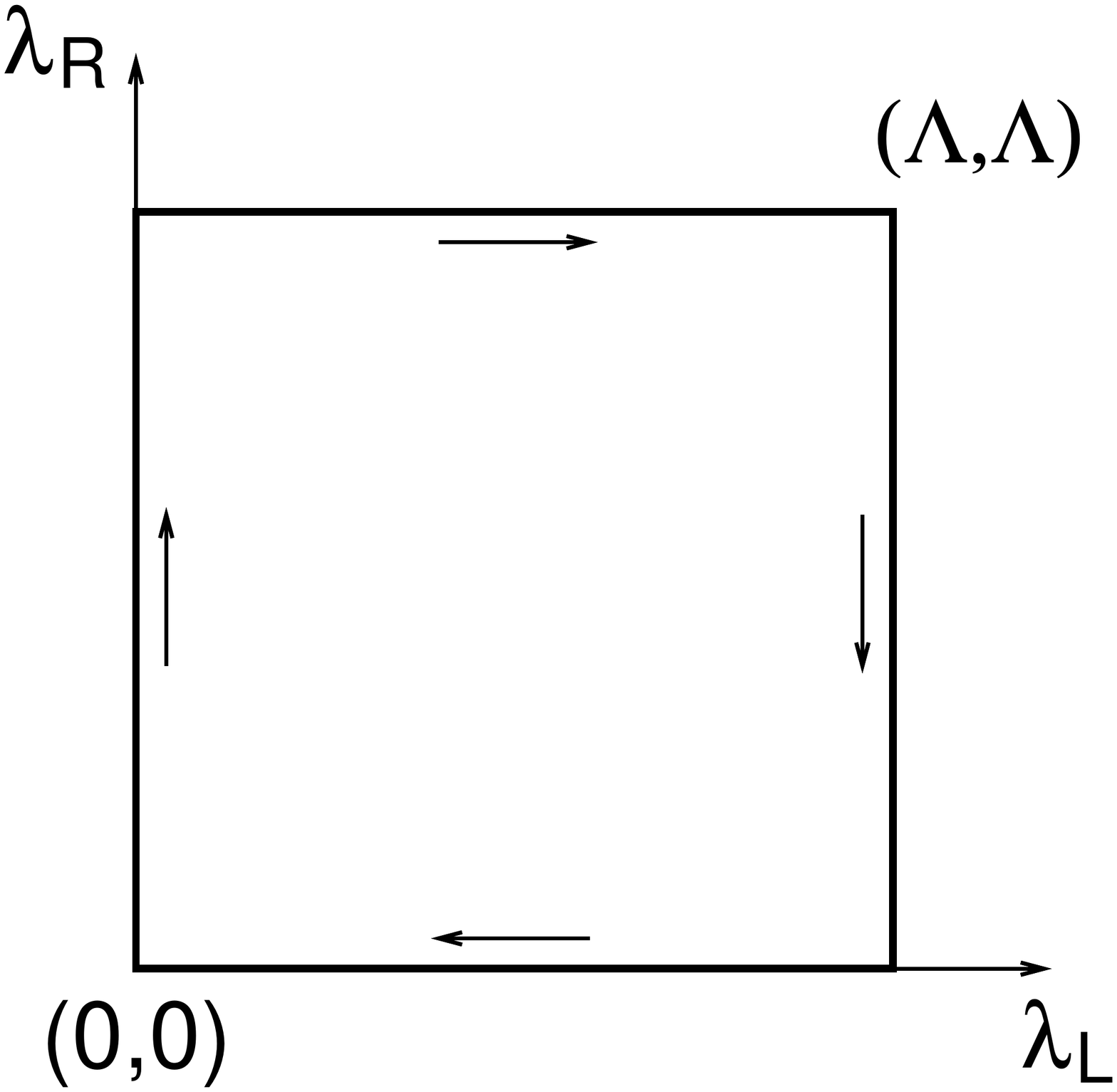}
   }
   \subfigure[\, Asymmetric protocol.]{
   \label{fig:asymmetricProtocol}
   \includegraphics[scale=0.25,angle=0]{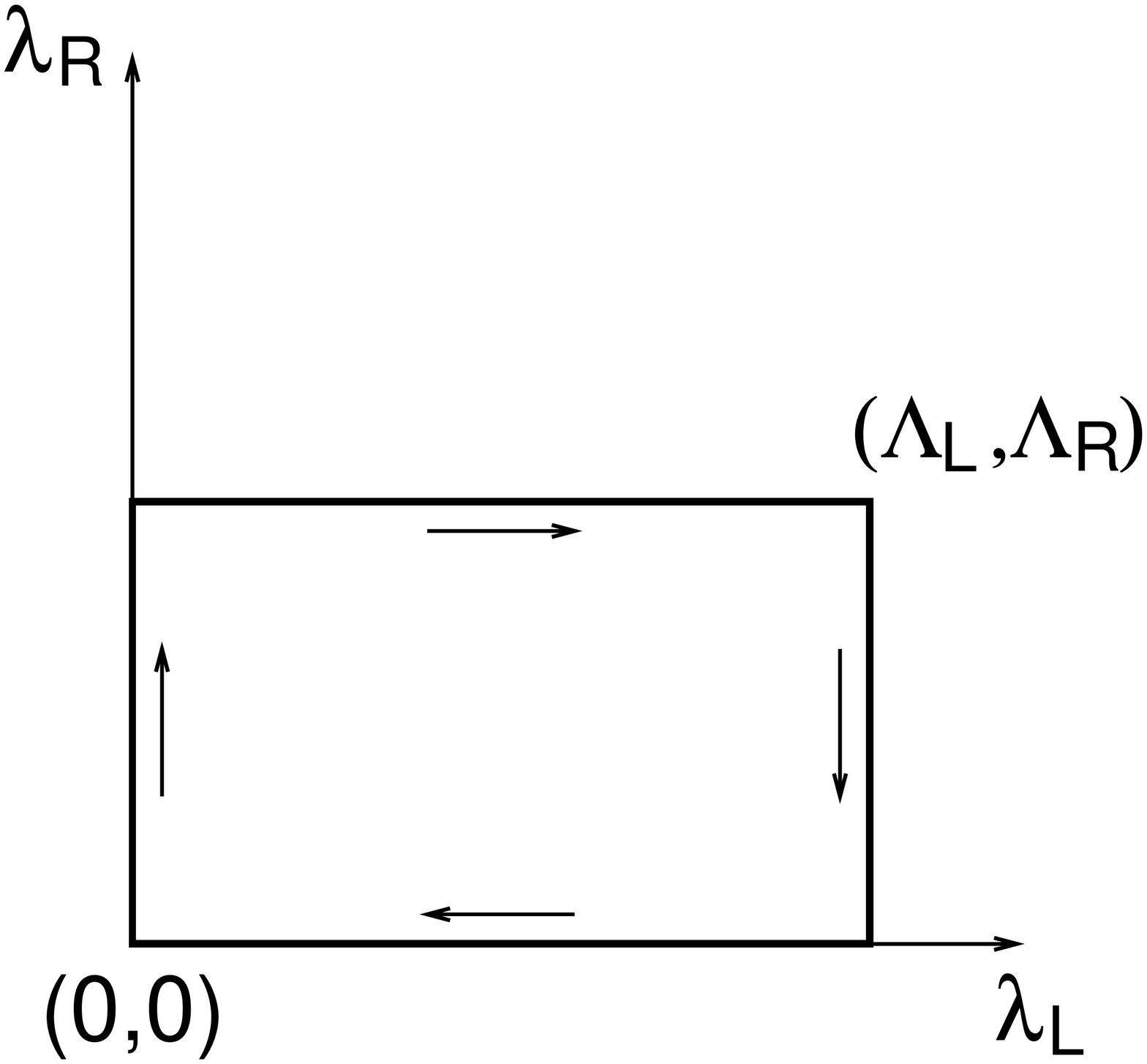}
   }
\caption{The cyclic protocols $\vec\lambda_c(t)$, depicted here, proceed clockwise from the origin.
}
\label{fig:protocol}
\end{figure}

\begin{figure}[tbp]
   \begin{center}
      \subfigure[$\,\,t=0$]{
         \label{fig:seq_t=0}
         \includegraphics[trim = 2in 1in 2in 0in , scale=0.18,angle=0]{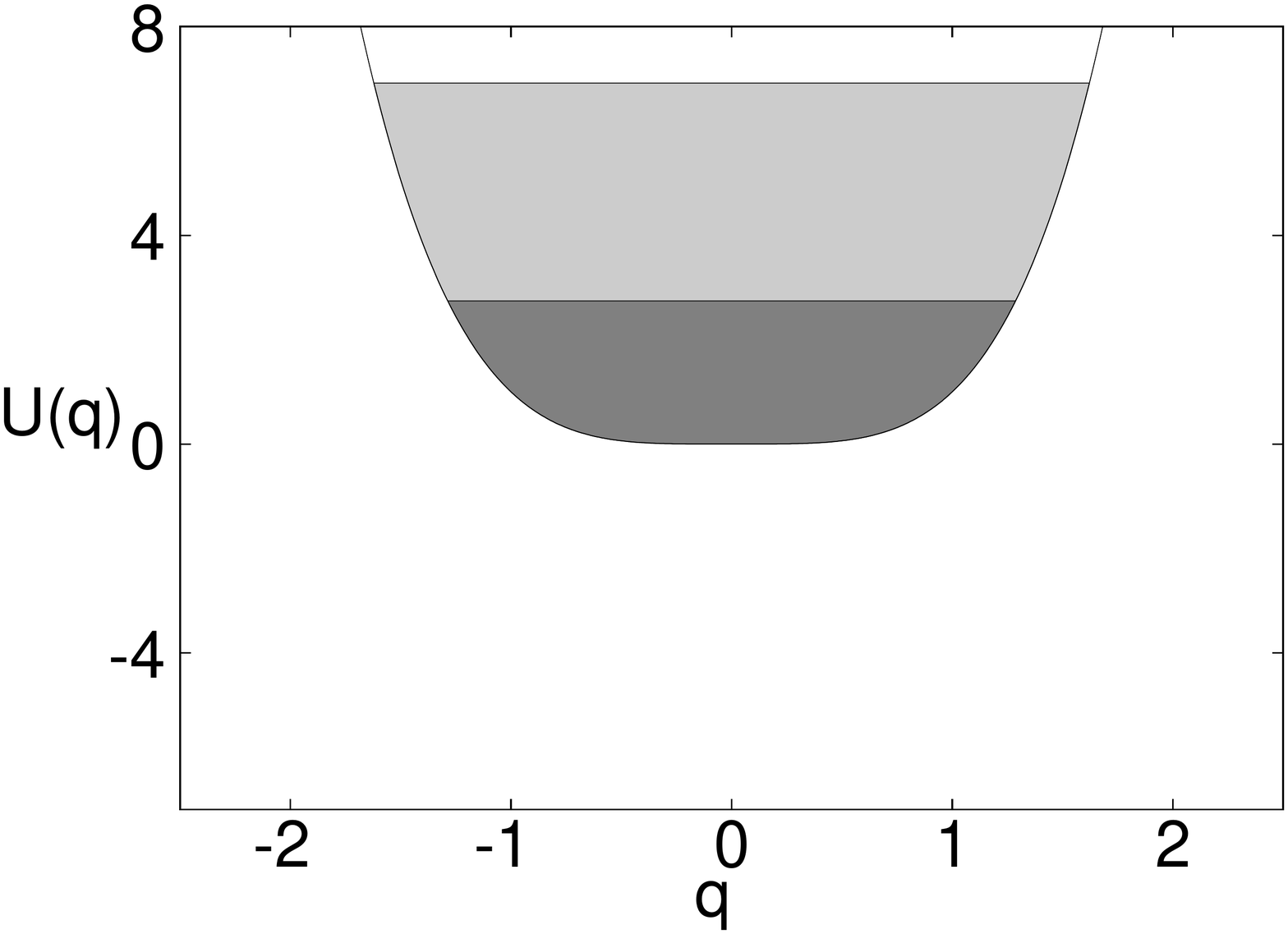} 
      }
      \subfigure[$\,\,t=\tau/8$]{
         \label{fig:seq_t=eighthtau}
         \includegraphics[trim = 0in 1in 2in 0in , scale=0.18,angle=0]{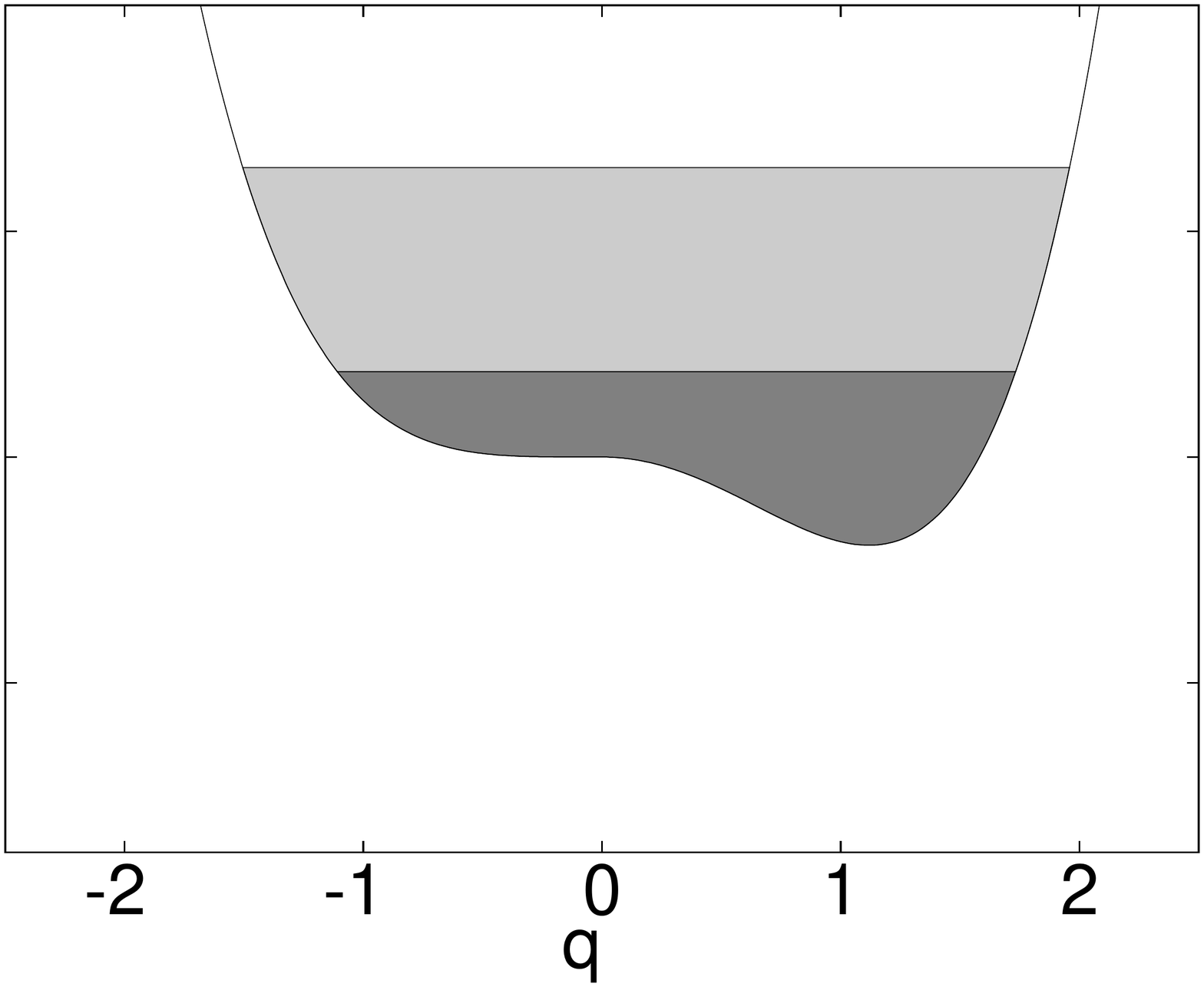} 
      } 
      \subfigure[$\,\,t=\tau/4$]{
         \label{fig:seq_t=quartertau}
         \includegraphics[trim = 0in 1in 2in 0in , scale=0.18,angle=0]{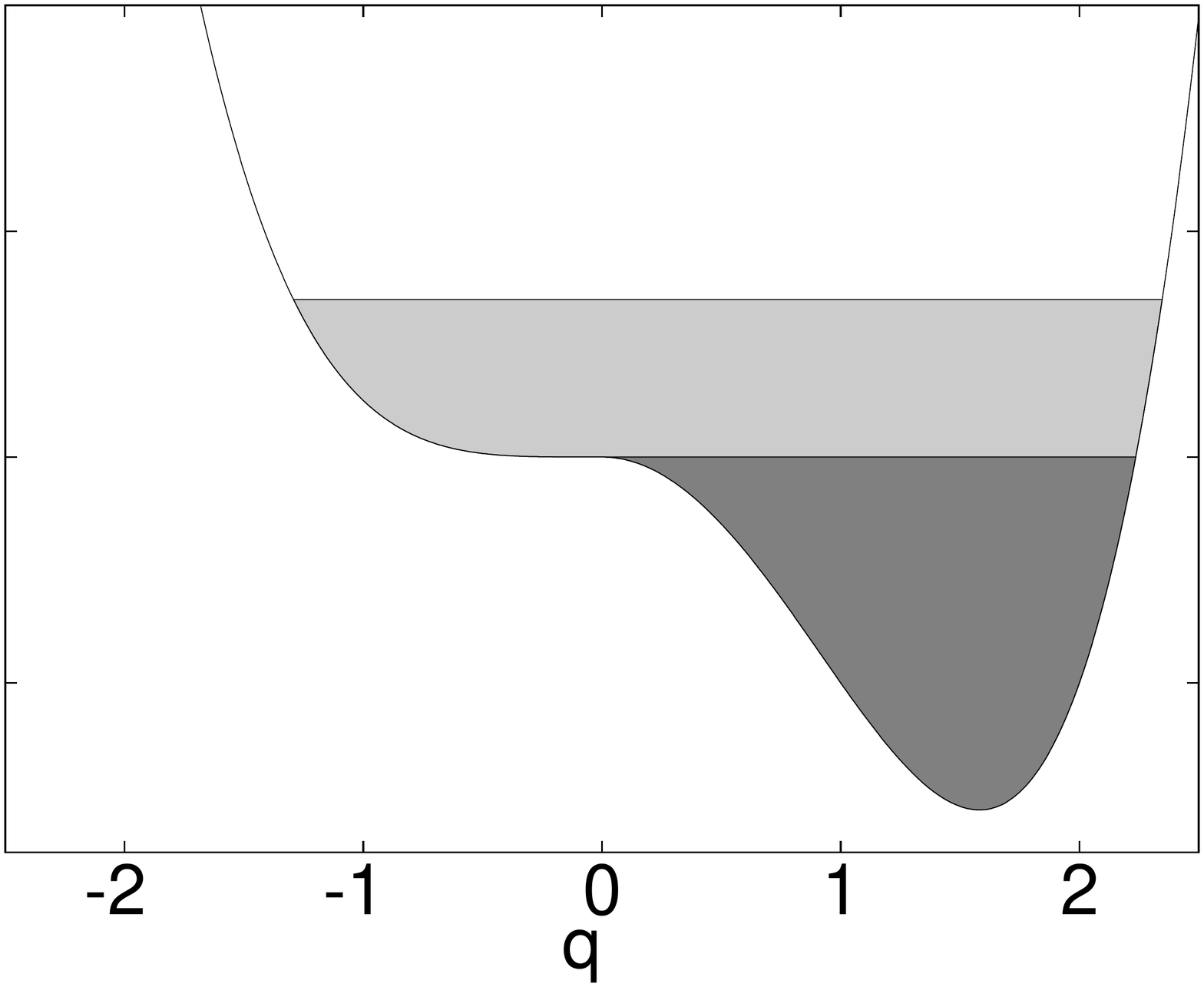} 
      } \\
      \subfigure[$\,\,t=\tau/2$]{
         \label{fig:seq_t=halftau}
         \includegraphics[trim = 2in 1in 2in 0in , scale=0.18,angle=0]{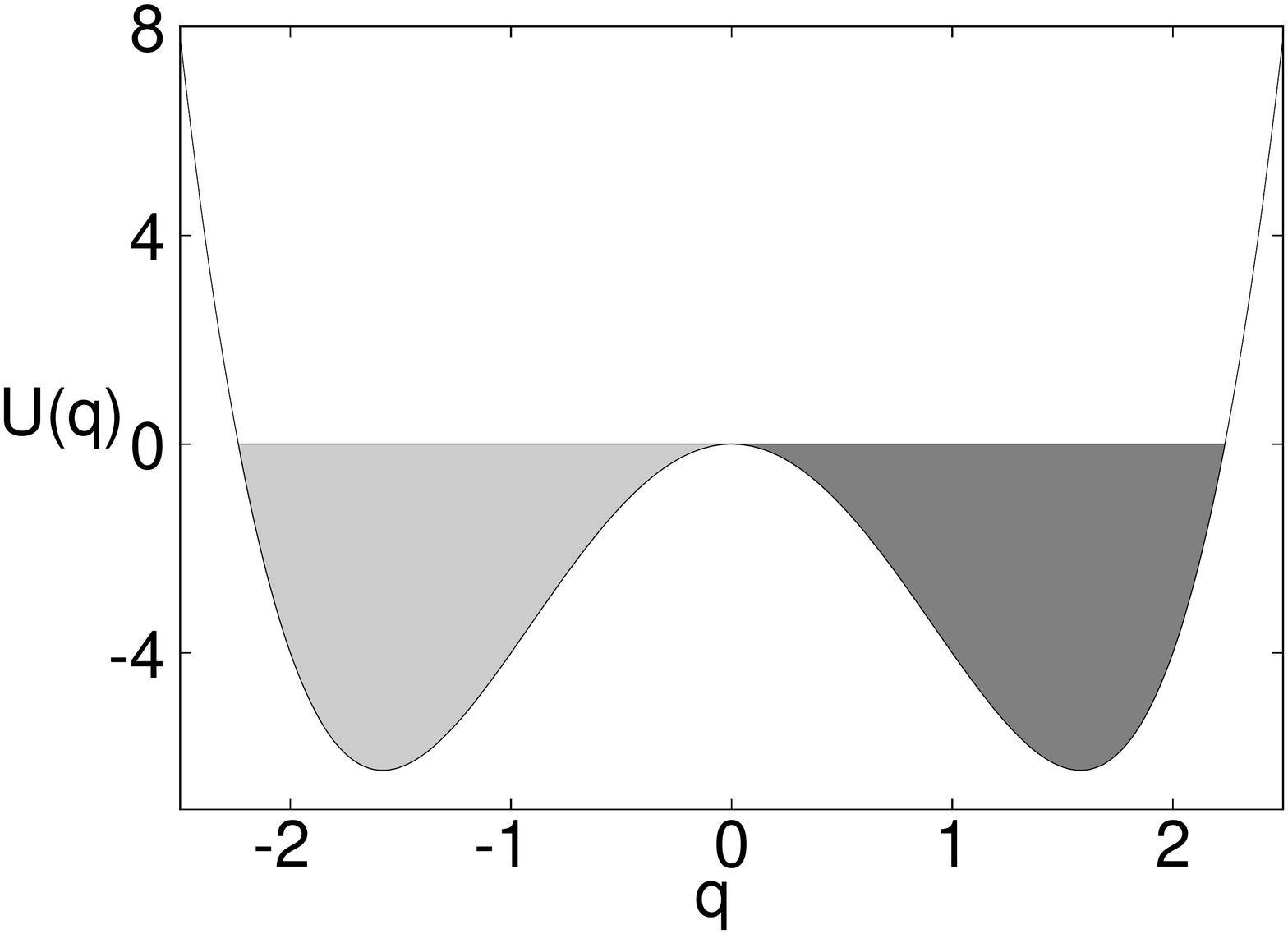} 
      } 
      \subfigure[$\,\,t=3\tau/4$]{
         \label{fig:seq_t=threequarterstau}
         \includegraphics[trim = 0in 1in 2in 0in , scale=0.18,angle=0]{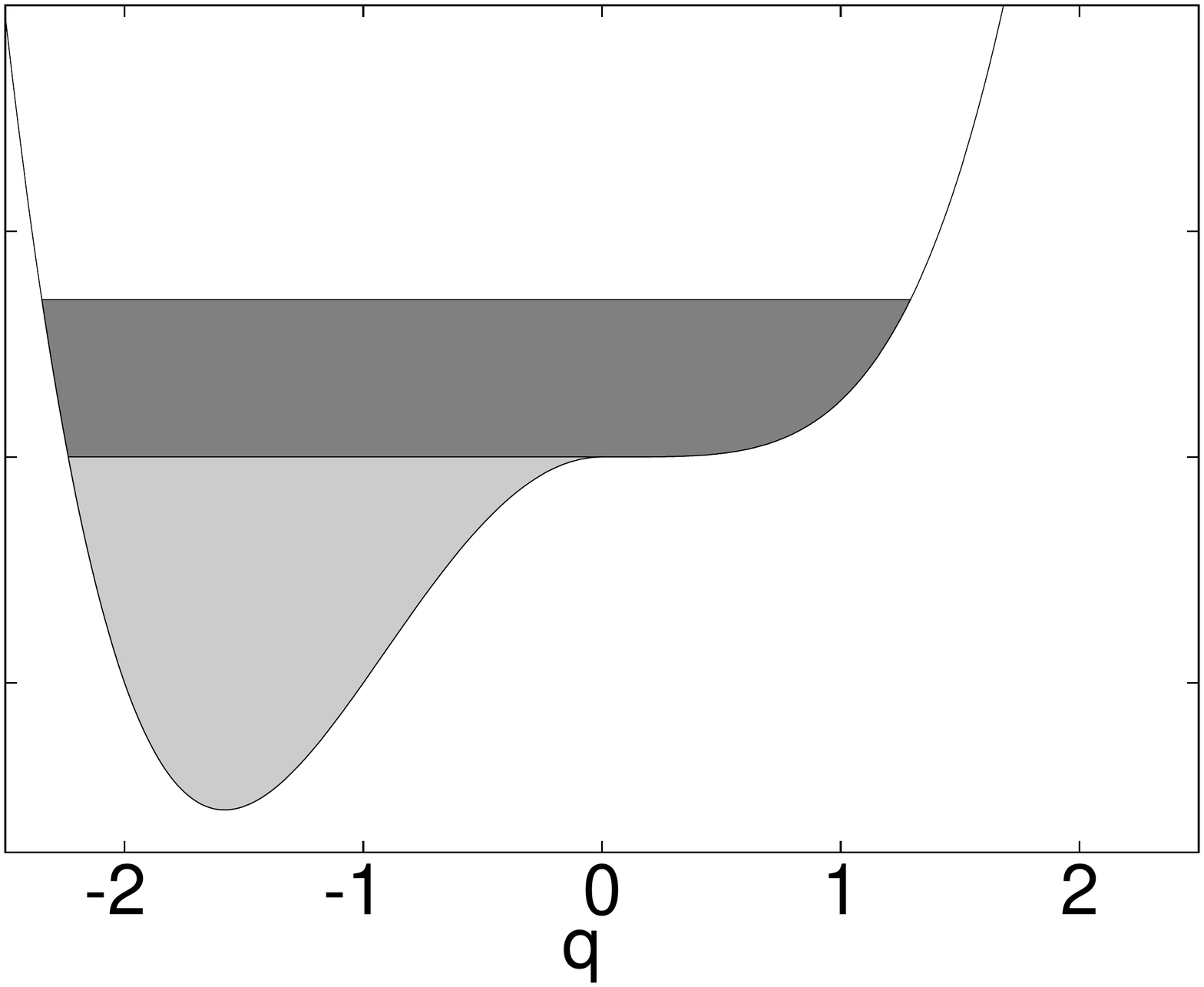} 
      } 
      \subfigure[$\,\,t=\tau$]{
         \label{fig:seq_t=tau}
         \includegraphics[trim = 0in 1in 2in 0in , scale=0.18,angle=0]{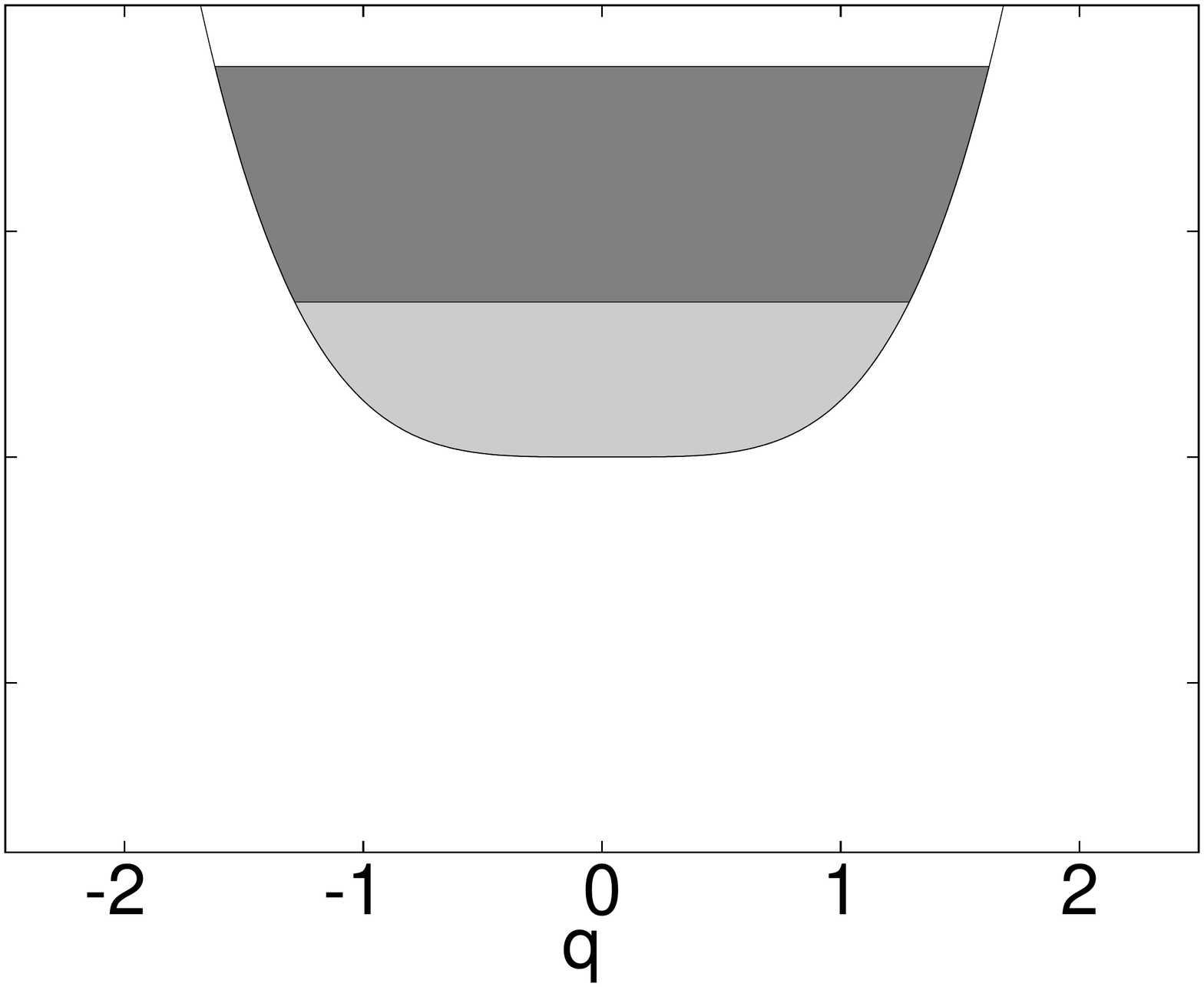} 
      }
   \end{center}
   \caption{Snapshots of the potential energy function as $\vec\lambda$ is varied according to the protocol shown in Fig.~\ref{fig:symmetricProtocol}, with $\Lambda = 5.0$ (hence $E_1 = 2.744$ and $E_2 = 6.914$, see Eq.~\ref{eq:defE1E2}).  
   The shaded regions illustrate the evolution of sets $I$ and $II$, in the quasi-static limit $\tau\rightarrow\infty$.}
    \label{fig:sequence}
\end{figure}

Now imagine a protocol $\vec\lambda_c(t)$ whereby the parameters are made to trace out the perimeter of the square shown in Fig.~
\ref{fig:symmetricProtocol}, starting and ending at $\vec\lambda=(0,0)$.
For simplicity we assume a constant speed,
$\vert {\rm d}\vec\lambda/{\rm d}t \vert = 4\Lambda/\tau$.
The deformation of the potential during this protocol can be pictured as follows.
Starting from the unperturbed quartic potential, the right well gradually drops down, forming a local minimum that moves from the origin to $\sqrt{\Lambda/2}$ (see Fig.~\ref{fig:seq_t=0} - \ref{fig:seq_t=quartertau}) as $\lambda_R$ increases from 0 to $\Lambda$.
Next, as $\lambda_L$ increases from 0 to $\Lambda$ the left well drops down, forming a local minimum that comes to rest at $-\sqrt{\Lambda/2}$, with a local maximum at the origin (Fig.~\ref{fig:seq_t=halftau}).
These two stages are then undone (Figs.~\ref{fig:seq_t=threequarterstau}, \ref{fig:seq_t=tau}).
The net effect is a piston-like pumping of the right and left wells.
For this protocol, let ${\bf z}_t$ denote a trajectory evolving under the time-dependent Hamiltonian $H({\bf z};\lambda_c(t))$.

For a given choice of $\Lambda$, let us define two energy values,
\begin{equation}
\label{eq:defE1E2}
E_1(\Lambda) = \left( \frac{1}{3I_0} \right)^{4/3} \, \Lambda^2
\qquad,\qquad
E_2(\Lambda) = \left( \frac{2}{3I_0} \right)^{4/3} \, \Lambda^2 ,
\end{equation}
where
\begin{equation}
\label{eq:I0}
I_0 = \int_{-1}^{+1} {\rm d}y \, \sqrt{ 1-y^4 }
= \frac{\sqrt{\pi}\,\Gamma(5/4)}{\Gamma(7/4)} \approx 1.74804 .
\end{equation}
These in turn define three regions of phase space, $I$, $II$, and $III$, according to the value of the unperturbed Hamiltonian $H_0({\bf z}) \equiv p^2/2 + q^4$:
\begin{equation}
\label{eq:regions}
\begin{split}
I \, &: 0 < H_0({\bf z}) < E_1 \\
II \, &: E_1 < H_0({\bf z}) < E_2 \\
III \, &: E_2 < H_0({\bf z})
\end{split}
\end{equation}
We now claim that when the protocol $\vec\lambda_c(t)$ shown in Fig.~\ref{fig:symmetricProtocol} is implemented quasi-statically, $\tau \rightarrow \infty$, then the net effect is to swap regions $I$ and $II$.
That is, trajectories with initial conditions ${\bf z}_0$ in region $I$ end with final conditions ${\bf z}_\tau$ in region $II$, and vice-versa.
(See, however, the discussion of subtleties associated with this limit, in Sec.~\ref{sec:discussion}.)
Fig.~\ref{fig:sequence} and the following paragraphs convey how this swap proceeds.
For convenience, we will use the terms {\it set $I$} and {\it set $II$} to refer to trajectories with initial conditions in regions $I$ and $II$ of phase space, respectively.
The shaded regions in Fig.~\ref{fig:sequence} depict the evolution of these sets of trajectories, as a sequence of snapshots from $t=0$ to $t=\tau$.

By Hamilton's equations we have
\begin{equation}
\label{eq:dHdt}
\frac{\rm d}{{\rm d}t}
H({\bf z}_t; \vec\lambda_t) =
\frac{{\rm d} \vec\lambda}{{\rm d}t} \cdot
\frac{\partial H}{\partial\vec\lambda} ({\bf z}_t;\vec\lambda_t) =
-q_t^2 \left[ \dot\lambda_L\, \theta(-q_t) + \dot\lambda_R\, \theta(+q_t) \right]
\end{equation}
where $\theta(\cdot)$ is the unit step function.
During the first stage of the process, $0 < t < \tau/4$, we have $\dot\lambda_L=0$ and $\dot\lambda_R > 0$, therefore as the right well drops down the value of $H({\bf z}_t;\vec\lambda_t)$ decreases whenever $q_t>0$.
As a result, some trajectories acquire negative energies ($H<0$) and become trapped in the right well.
As shown in Fig.~\ref{fig:seq_t=quartertau} -- and as justified quantitatively by Eqs.~\ref{eq:OmegaEL} - \ref{eq:justification} below -- at the end of this stage the trajectories belonging to set $I$ are trapped.

During the second stage, $\tau/4 < t < \tau/2$, the left well drops down, trapping the trajectories in set $II$.
As this occurs, the trajectories in set $I$ remain trapped in the right well.

From $\tau/2 < t < 3\tau/4$, as the right well rises and ultimately disappears, the trajectories in set $I$ gain energy (Fig.~\ref{fig:seq_t=threequarterstau}), and during the fourth and final stage, $3\tau/4 < t < \tau$, all trajectories gain energy as the left well gradually rises until it disappears.
The situation at $t=\tau$, shown in Fig.~\ref{fig:seq_t=tau}, reflects the swap that has occurred between sets $I$ and $II$, relative to Fig.~\ref{fig:seq_t=0}.

Due to adiabatic averaging, the energy-ordering of the trajectories within each set remains fixed in the quasi-static limit: if we were to subdivide the lightly shaded region $II$ in Fig.~\ref{fig:seq_t=0} into a stack of narrow horizontal bands, then the vertical ordering of these bands would remain unchanged throughout the process.

A proper analysis of this process involves the theory of adiabatic invariants, with careful attention paid to the phase space separatrix that is present during the interval $\tau/4 < t < 3\tau/4$, when $U(q)$ has a local maximum at $q=0$~\cite{Tennyson1986,Cary1986}.
However, the essence of what occurs should be intuitively clear from the above discussion.
A useful analogy is provided by imagining a container initially filled with three layers of a viscous, incompressible fluid, labeled $I$, $II$ and $III$ in vertically ascending order.
Two syringes are attached to the bottom of the container.
First one syringe extracts the lowest layer $I$ of the fluid, bringing layer $II$ to the bottom of the container.
Next, the other syringe extracts layer $II$.
Then the fluid layers are re-injected in the same order in which they were removed, resulting in the rearrangement of these layers.

The incompressibility of the fluid in this analogy corresponds to Liouville's theorem: phase space volume is preserved under Hamiltonian dynamics.
To justify quantitatively our assertion that the protocol $\vec\lambda_c(t)$ swaps regions $I$ and $II$, we must show that the phase space volumes corresponding to the darkly shaded regions in Figs.~\ref{fig:seq_t=0} and Figs.~\ref{fig:seq_t=halftau} are equal (in other words, it is precisely the trajectories in set $I$ that get trapped in the right well), and similarly that the phase space volumes of the lightly shaded regions in Figs.~\ref{fig:seq_t=0} and Figs.~\ref{fig:seq_t=halftau} are equal.

Let $\Omega(E;\vec\lambda)$ denote the volume of phase space enclosed by the surface $H({\bf z};\vec\lambda) = E$:
\begin{equation}
\begin{split}
\label{eq:OmegaEL}
\Omega(E;\vec\lambda) &= \int {\rm d}{\bf z} \ \theta \left[ E - H({\bf z};\vec\lambda) \right] \\
&= \int_{E>U} {\rm d} q \, \sqrt{8 \left[ E - U(q;\vec\lambda) \right]}
\end{split}
\end{equation}
where we have integrated over momentum to get to the second line.
When either $E=0$ or $\vec\lambda = \vec 0$ the remaining integral can be evaluated analytically:
\begin{subequations}
\begin{eqnarray}
\label{eq:OmegaE0}
\Omega(E;\vec 0) &=& \int_{-E^{1/4}}^{+E^{1/4}} {\rm d}q \, \sqrt{ 8 (E-q^4) } = \sqrt{8} E^{3/4} I_0 \\
\label{eq:Omega0L}
\Omega(0;\vec\lambda) &=& \int_{-\sqrt{\lambda_L}}^{\sqrt{\lambda_R}} {\rm d} q \, \sqrt{ -8 U(q;\vec\lambda) }
= \sqrt{\frac{8}{9}} \left( \lambda_L^{3/2} + \lambda_R^{3/2} \right)
= \Omega_L + \Omega_R
\end{eqnarray}
\end{subequations}
with $I_0$ given by Eq.~\ref{eq:I0}.
The quantity
\begin{equation}
\label{eq:OmegaL_def}
\Omega_L(\lambda_L) \equiv \sqrt{\frac{8}{9}} \, \lambda_L^{3/2}
\end{equation}
is the volume of phase space for which $H<0$ and $q<0$, and $\Omega_R(\lambda_R)$ is defined similarly for $H<0$ and $q>0$.

Using Eq.~\ref{eq:OmegaE0}, the phase space volumes of regions $I$ and $II$, defined by Eq.~\ref{eq:regions}, are
\begin{equation}
\label{eq:omega_I_II}
\Omega_I = \sqrt{8} E_1^{3/4} I_0
\qquad,\qquad
\Omega_{II} = \sqrt{8} \left( E_2^{3/4} - E_1^{3/4} \right) I_0
\end{equation}
In Fig.~\ref{fig:seq_t=halftau} the lightly and darkly shaded regions  correspond to phase space volumes $\Omega_L(\Lambda)$ and $\Omega_R(\Lambda)$, respectively, which are equal in value:
\begin{equation}
\Omega_L(\Lambda) = \Omega_R(\Lambda) = \sqrt{\frac{8}{9}} \, \Lambda^{3/2}
\end{equation}
Combining these results with Eq.~\ref{eq:defE1E2} we find that
\begin{equation}
\label{eq:justification}
\Omega_I = \Omega_R(\Lambda)
\qquad,\qquad
\Omega_{II} = \Omega_L(\Lambda)
\end{equation}
This establishes that our qualitative description of what occurs during this process, as illustrated in Fig.~\ref{fig:sequence}, is indeed consistent with the preservation of phase space volume, as mandated by Liouville's theorem.

The picture developed in the preceding paragraphs suggests the following relationship between the initial ($E_i$) and final ($E_f$) energy of the system, in the limit $\tau\rightarrow\infty$:
\begin{subequations}
\label{eq:phaseSpaceSwap}
\begin{alignat}{2}
\Omega(E_f ; \vec 0) &= \Omega(E_i ; \vec 0) + \Omega_{II}  &\qquad &\text{if $\quad 0 < E_i < E_1$} \\
\Omega(E_f ; \vec 0) &= \Omega(E_i ; \vec 0) - \Omega_{I}   & &\text{if $\quad E_1 < E_i < E_2$} \\
\Omega(E_f ; \vec 0) &= \Omega(E_i ; \vec 0) & &\text{if $\quad E_2 < E_i$}
\end{alignat}
\end{subequations}
with $E_1$ and $E_2 = 2^{4/3}E_1$ determined by the value of $\Lambda$ (Eq.~\ref{eq:defE1E2}).
Combining these results with Eq.~\ref{eq:omega_I_II} (note that $\Omega_I=\Omega_{II}$) we obtain
\begin{equation}
\label{eq:Ef}
E_f =
\begin{cases}
\left( E_i^{3/4} + E_1^{3/4} \right)^{4/3} & \text{if $\quad 0 < E_i < E_1$} \\
\left( E_i^{3/4} - E_1^{3/4} \right)^{4/3} & \text{if $\quad E_1 < E_i < E_2$} \\
\qquad E_i & \text{if $\quad E_2 < E_i$} 
\end{cases}
\end{equation}

As a test of Eq.~\ref{eq:Ef}, we sampled $10^5$ initial conditions ${\bf z}_0 = (q_0,p_0)$ from a microcanonical ensemble at energy $E_i = H_0({\bf z}_0) = 2.8$, near the bottom of region $II$ (see Fig.~\ref{fig:sequence}).
For each initial condition ${\bf z}_0$ we generated a trajectory ${\bf z}_t$ by integrating Hamilton's equations as the parameters were varied as in Fig.~\ref{fig:symmetricProtocol}, with $\tau = 12000$.
The resulting distribution of final energies $E_f = H_0({\bf z}_\tau)$, spanning a range from $E_{f,{\rm min}}= 0.0030$ to $E_{f,{\rm max}}= 0.0150$, was characterized by a mean value $\overline{E_f} = 0.0106$ and a standard deviation $\sigma_{E_f} = 0.0014$, in excellent agreement with the value $E_f = 0.0104$ predicted by Eq.~\ref{eq:Ef}.
(The small discrepancies reflect the fact that the duration $\tau = 12000$ is finite.)
While these numerical results support the analysis leading to Eq.~\ref{eq:Ef}, some caveats are in order.
In particular, Liouville's theorem itself rules out the possibility that {\it all} initial conditions with energy $E_i=2.8$ lead to a net decrease of energy, $E_f<E_i$.
We defer a discussion of this issue to Sec.~\ref{sec:discussion}.

\begin{figure}[tbp]
   \begin{center}
      \subfigure[$\,\,t=0$]{
         \includegraphics[trim = 2in 1in 2in 0in , scale=0.18,angle=0]{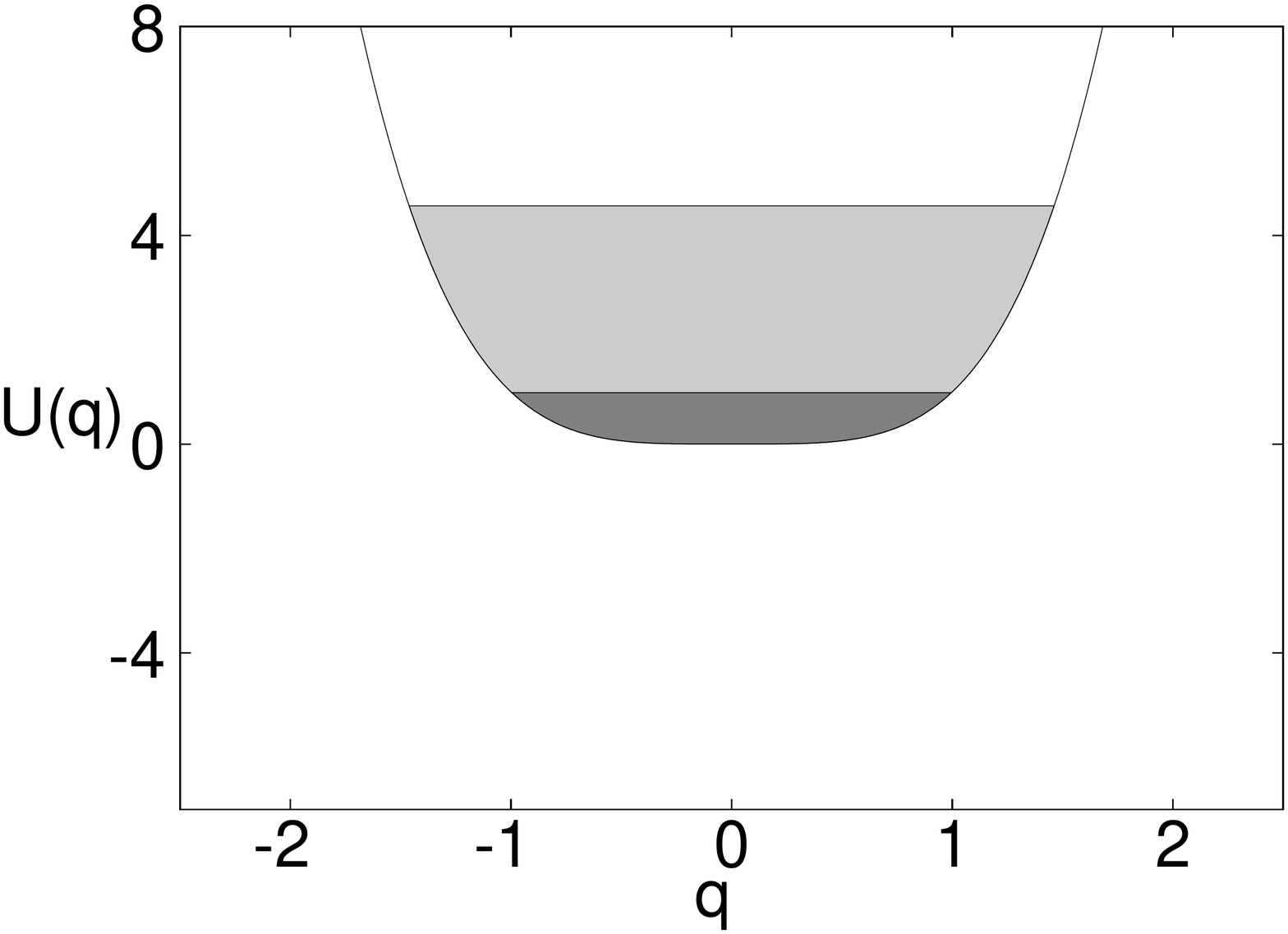} 
      }
      \subfigure[$\,\,t=\tau/2$]{
         \includegraphics[trim = 0in 1in 2in 0in , scale=0.18,angle=0]{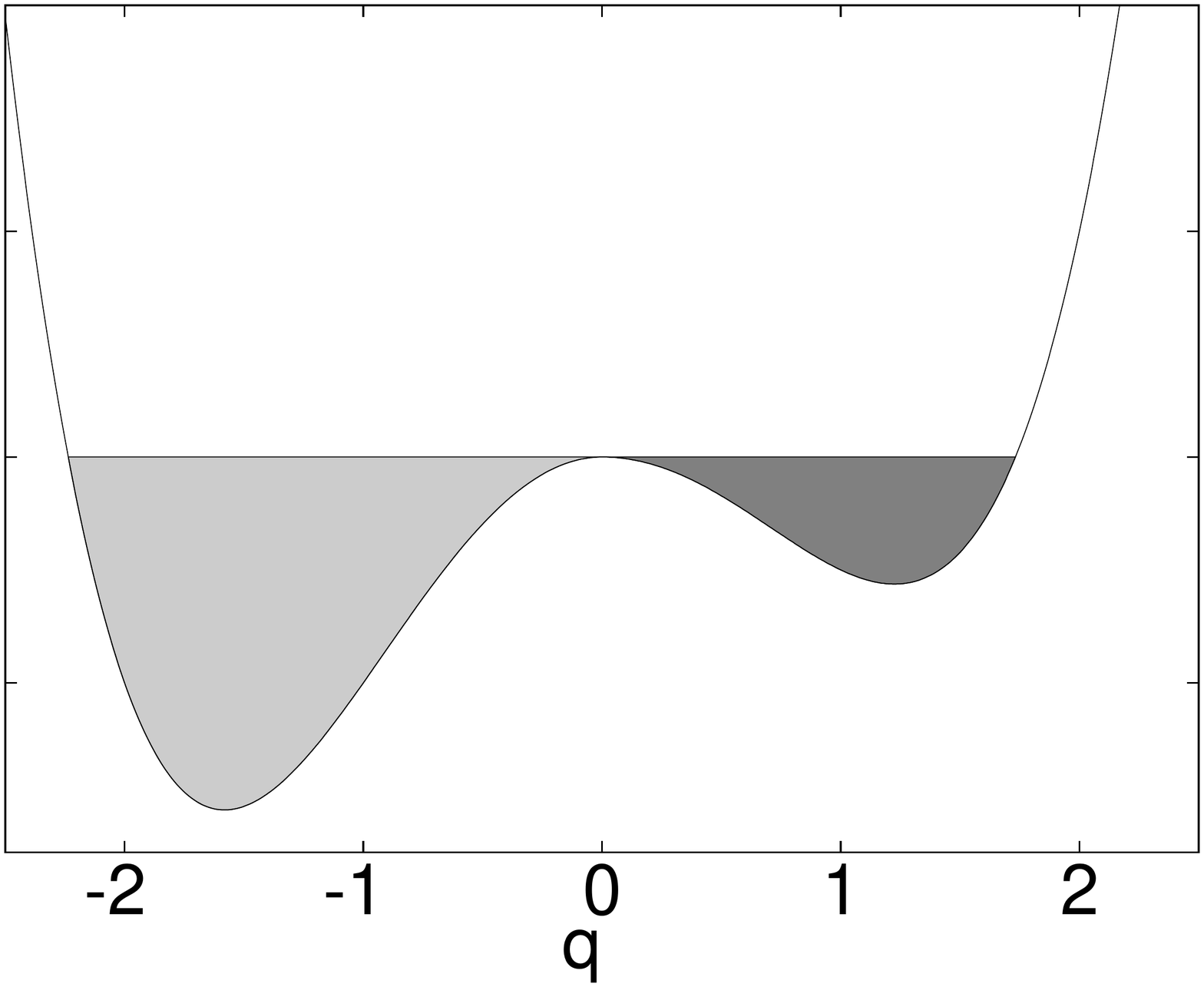} 
      } 
      \subfigure[$\,\,t=\tau$]{
         \includegraphics[trim = 0in 1in 2in 0in , scale=0.18,angle=0]{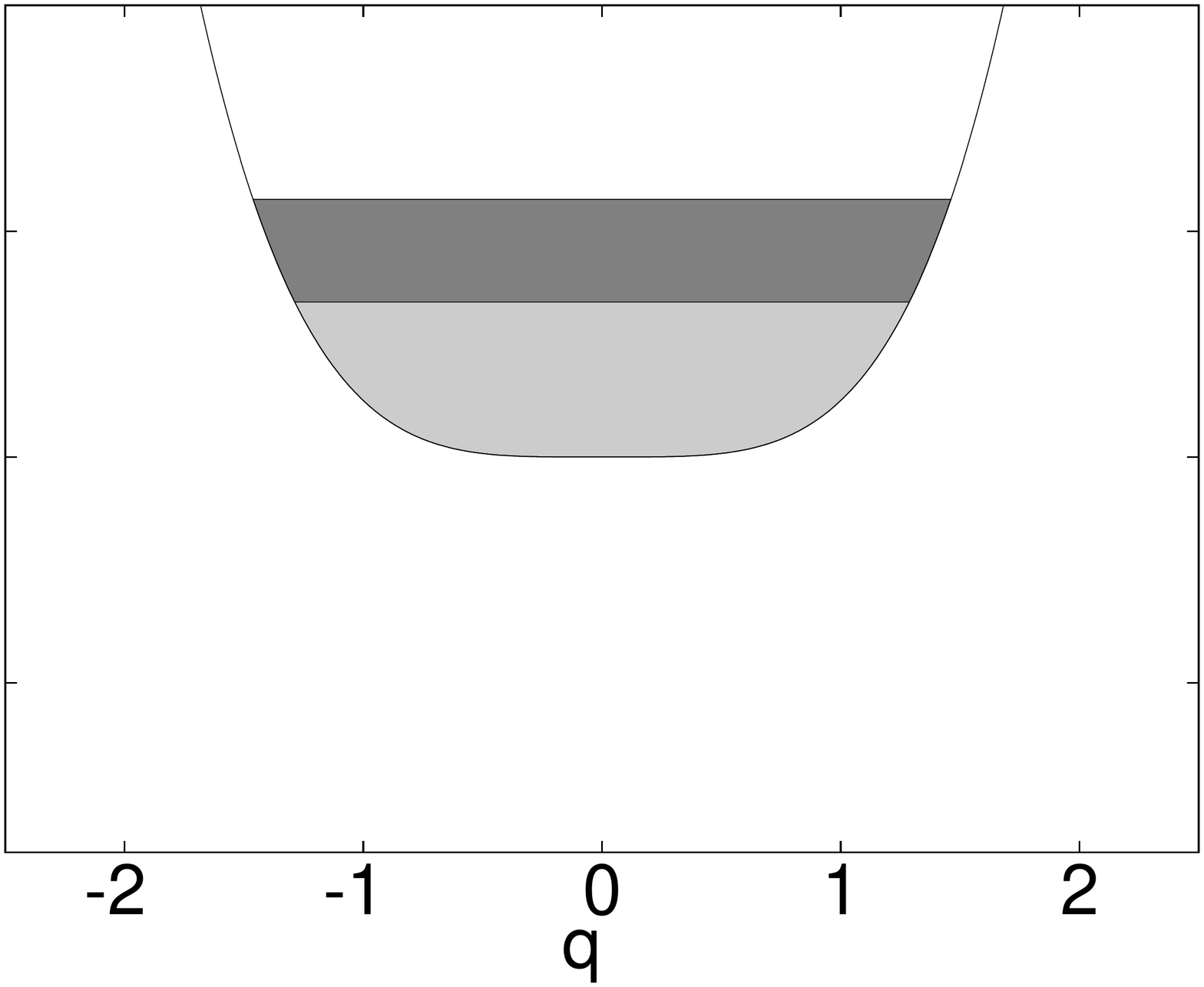} 
      }
   \end{center}
   \caption{Similar to Fig.~\ref{fig:sequence}, but for an asymmetric protocol, Fig.~\ref{fig:asymmetricProtocol}, with $\Lambda_L = 5.0$ and $\Lambda_R = 3.0$.
   The phase space volume of set $I$ (the darkly shaded region) remains constant, as does the volume of set $II$ (lightly shaded), but the two volumes differ: $\Omega_I \ne \Omega_{II}$.}
   \label{fig:asymmetricSequence}
\end{figure}

To this point we have considered a symmetric protocol, Fig.~\ref{fig:symmetricProtocol}, in which each well reaches the same maximal depth, determined by the value of $\Lambda$ (Fig.~\ref{fig:seq_t=halftau}).
However, the analysis is easily generalized to the asymmetric protocol shown in Fig.~\ref{fig:asymmetricProtocol}, in which the parameters are varied around a rectangle with corners at $(0,0)$ and $(\Lambda_L,\Lambda_R)$.
Regions $I$, $II$ and $III$ are defined as in Eq.~\ref{eq:regions}, but now the energies $E_1$ and $E_2$ are defined by
\begin{equation}
\label{eq:E1E2_asymm}
E_1(\vec\Lambda) = \left( \frac{\Lambda_R^{3/2}}{3I_0} \right)^{4/3}
\qquad,\qquad
E_2(\vec\Lambda) = \left( \frac{\Lambda_R^{3/2} + \Lambda_L^{3/2}}{3I_0} \right)^{4/3} 
\end{equation}
When the protocol is implemented quasi-statically, the net result is a rearrangement of sets $I$ and $II$, as depicted in Fig.~\ref{fig:asymmetricSequence}.
Eq.~\ref{eq:phaseSpaceSwap} now leads to the result
\begin{equation}
\label{eq:Ef_asymm}
E_f =
\begin{cases}
\left( E_i^{3/4} + E_2^{3/4} - E_1^{3/4} \right)^{4/3} & \text{if $0 < E_i < E_1$} \\
\left( E_i^{3/4} - E_1^{3/4} \right)^{4/3} & \text{if $E_1 < E_i < E_2$} \\
\qquad E_i & \text{if $E_2 < E_i$}
\end{cases}
\end{equation}
The viscous fluid analogy also applies to this situation, only now the syringes remove different quantities of fluid, $\Omega_I \ne \Omega_{II}$.
Alternatively, the processes illustrated in Figs.~\ref{fig:sequence} and \ref{fig:asymmetricSequence} are analogous to a simple shuffle of a deck of cards, in which a stack of adjacent cards (region $II$) is removed from the middle of the deck and transferred to the bottom.

It should now be clear how to design a quasi-static protocol that lowers the energy of the system almost to zero, for a given initial energy $E_i = H_0({\bf z}_0)$.
Namely, we choose $\Lambda_R$ such that $E_i$ is slightly above $E_1$, thus locating the initial conditions near the bottom of region $II$.
If we then implement the protocol shown in Fig.~\ref{fig:protocol}, in either its symmetric ($\Lambda_L = \Lambda_R = \Lambda$) or asymmetric ($\Lambda_L \ne \Lambda_R$) version, the system will be trapped near the bottom of the left well at $t=\tau/2$, and will end the process with $E_f \approx 0$.
This outcome is independent of the value of $\Lambda_L$, which simply determines the width (in energy) of region $II$.

\section{Exorcising Maxwell's Demon}
\label{sec:erasure}

Let us now return to the perpetual-motion device of the second kind proposed in the Introduction:
after equilibrating the system with a thermal reservoir at temperature $T$ (step 1), we measure the initial energy $E_i$ (step 2), then choose a protocol that reduces the energy near to zero (step 3).
The amount of work we extract during this cycle -- equivalently, {\it minus} the amount of work we perform on the system -- is given by
\begin{equation}
W_{\rm extracted} = 
-W = E_i - E_f  < E_i
\end{equation}
If we repeat this process many times, then the average work extracted per cycle satisfies
\begin{equation}
\label{eq:threefourths}
\langle W_{\rm extracted} \rangle < \langle E_i \rangle = \int {\rm d}{\bf z}_0 \, p^{\rm eq}({\bf z}_0) \, H_0({\bf z}_0)
= \frac{3}{4} \, \beta^{-1}
\qquad
(\beta^{-1} \equiv k_BT)
\end{equation}
where the canonical distribution $p^{\rm eq} \propto \exp( -\beta H_0 )$ reflects initial equilibration with the reservoir.\footnote{
In Eq.~\ref{eq:threefourths} we have used the identity
$\langle E \rangle = -(\partial/\partial\beta) \ln Z$,
with $Z \equiv \int {\rm d}{\bf z} \exp(-\beta H_0) = \sqrt{8\pi} \Gamma(5/4) \beta^{-3/4}$.
}
To approach this upper bound of $(3/4)k_BT$ per cycle, in which the thermal energy of the system is entirely converted to work ($E_f = 0$),
the initial energy must be measured with high precision, allowing us to choose a protocol for which $E_i - E_1(\vec\Lambda)$ is tiny but positive (Eqs.~\ref{eq:Ef}, \ref{eq:Ef_asymm}).
However, as mentioned in the Introduction, these measurements generate information that must ultimately be erased, at a cost of $\beta^{-1} \ln 2$ per bit.
There is a competition at play here: increased precision brings us closer to the maximal extracted work, but carries the penalty of increased accumulation of information.

To address this issue, imagine a measurement apparatus that reports the initial energy of the system with finite precision.
Specifically, given the initial microstate ${\bf z}_0$, the apparatus outputs one of $K$ values associated with specified energy intervals $A$, $B$, $C$, $\cdots$.
Taking $K=4$ for purpose of illustration, the apparatus outputs $A$, $B$, $C$, or $D$ according to
\begin{equation}
\label{eq:measure_K=4}
\begin{split}
A \, &: 0 < H_0({\bf z}_0) < E_A \\
B\, &: E_A < H_0({\bf z}_0) < E_B \\
C\, &: E_B < H_0({\bf z}_0) < E_C \\
D \, &: E_C < H_0({\bf z}_0)
\end{split}
\end{equation}
where the values $E_A$, $E_B$, and $E_C$ are fixed properties of the apparatus.

Now consider the following strategy for choosing a cyclic protocol, based on the output of the measurement apparatus.

\noindent $\bullet$
\underline{$\textrm{Output}=A$}:
Do nothing to the system, as it is already in the lowest-energy interval.

\noindent $\bullet$
\underline{$\textrm{Output}=B$}:
Using Eq.~\ref{eq:E1E2_asymm}, set $E_1(\vec\Lambda) = E_A$ and $E_2(\vec\Lambda) = E_B$, that is choose $(\Lambda_L,\Lambda_R)$ so that interval $B$ in Eq.~\ref{eq:measure_K=4} corresponds to region $II$ in Eq.~\ref{eq:regions}.
Next, implement the asymmetric protocol of Fig.~\ref{fig:asymmetricProtocol}, under which initial conditions from this region are transferred to the bottom of the potential well, as in Fig.~\ref{fig:asymmetricSequence}.

\noindent $\bullet$
\underline{$\textrm{Output}=C$}:
Set $E_1(\vec\Lambda) = E_B$ and $E_2(\vec\Lambda) = E_C$, then implement the asymmetric protocol.
Again, the energy interval containing the initial conditions -- interval $C$, in this case -- is shuffled to the bottom of the potential.

\noindent $\bullet$
\underline{$\textrm{Output}=D$}:
Set $E_1(\vec\Lambda) = E_C$ and $E_2(\vec\Lambda) = E^*$, where $E^* > E_C$ is an arbitrary cutoff energy, then implement the asymmetric protocol.
In this case, initial conditions from the region between $E_C$ and $E^*$ are transferred to the bottom of the potential, whereas if $H({\bf z}_0)>E^*$ the protocol produces no net change in the energy of the system.

This strategy takes advantage of the limited knowledge provided by the measurement of the initial energy.
When it is implemented, the energy of the system decreases (that is, $E_f<E_i$) if $E_A<E_i<E^*$, and remains unchanged otherwise.
Thus, on average per cycle, work is extracted from the system,
\begin{equation}
\langle W_{\rm extracted} \rangle > 0
\end{equation}
and ultimately from the reservoir that replenishes the system's energy.

Over $N\gg 1$ repetitions of the process, the measurement apparatus generates a symbolic string of length $N$, of the form $BDCCADA\cdots$.
Letting $P_X$ denote the probability of outcome $X \in \{A,B,C, D\}$ in a given measurement, the number of bits required to encode this string is given by
\begin{equation}
\label{eq:Nbits}
N_{\rm bits} = N {\cal H}/\ln 2,
\end{equation}
where 
\begin{equation}
\label{eq:shannon}
{\cal H}=-\sum_X P_X \ln P_X
\end{equation}
is the Shannon entropy of the measurement~\cite{Cover2006}.
Now, both $\langle W_{\rm extracted}\rangle$ and ${\cal H}$ depend on $E_A$, $E_B$ and $E_C$, and the former also depends on $E^*$.
In the following section we establish that, no matter what values these parameters take, the inequality
\begin{equation}
\label{eq:inequality}
\langle W_{\rm extracted}\rangle \le \beta^{-1} {\cal H}
\end{equation}
is satisfied.
The extraction of work thus comes at the cost of the accumulation of information: on average, at least one bit is written per $\beta^{-1} \ln 2$ of extracted work.~\footnote{
In the original Szilard engine, which involves a single particle in a chamber, this relationship is straightforward: the determination whether the particle is in the left or right half of the chamber produces exactly one bit of information, ${\cal H} = \ln 2$, and standard thermodynamics gives the amount of work extracted during the subsequent isothermal expansion, $W_{\rm extracted}=\beta^{-1} \ln 2$.
}

We now turn our attention to the eventual cost of erasing this information.
By Landauer's principle, the average work required to erase one bit of information is no less than $\beta^{-1} \ln 2$.
Therefore, since the number of bits generated per cycle is ${\cal H}/\ln 2$ (Eq.~\ref{eq:Nbits}), the average work required to erase the information accumulated in one cycle of operation satisfies
\begin{equation}
\label{eq:W_erasure}
\langle W_{\rm erasure} \rangle \ge \beta^{-1} {\cal H}
\end{equation}

Combining Eqs.~\ref{eq:inequality} and \ref{eq:W_erasure}, we find that the work required to erase the accumulated information exceeds -- or at best, matches -- the work extracted during the cycle:
\begin{equation}
\label{eq:ineq_chain}
\langle W_{\rm extracted} \rangle \le \beta^{-1} {\cal H} \le \langle W_{\rm erasure}\rangle
\end{equation}
Thus our model obeys the Kelvin-Planck statement of the second law, as it had better do!
Eq.~\ref{eq:ineq_chain} highlights the two logically distinct steps we take in reconciling our model with the second law.
Although the second half of this inequality chain (that is, Landauer's principle) is derived by appeal to the second law itself~\cite{Landauer1961}, the first half (Eq.~\ref{eq:inequality}) is obtained without assuming the second law:
in Sec.~\ref{subsec:bound_on_work} we do not infer Eq.~\ref{eq:inequality} by arguing that the second law demands it, rather we will derive this inequality directly.

Eq.~\ref{eq:inequality} is a special case of an inequality recently derived by 
Sagawa and Ueda (see Eq.~3 of Ref.~\cite{Sagawa2010} or, in the quantum setting, Eq.~14 of Ref.~\cite{Sagawa2008}), which generalizes the second law of thermodynamics to processes with feedback, such as the one considered in this paper.
This inequality also follows readily from recent generalizations~\cite{Sagawa2010,Ponmurugan2010,Horowitz2010} of the nonequilibrium work relation~\cite{Jarzynski1997a} and Crooks's fluctuation theorem~\cite{Crooks1999} to nonequilibrium processes with feedback.
In the following derivation, we do not directly invoke these results, instead we provide a self-contained analysis that is pertinent to our particular model.

\subsection{Bound on work}
\label{subsec:bound_on_work}

Consider a cyclic process with the measurement apparatus described by Eq.~\ref{eq:measure_K=4} above.
For initial conditions ${\bf z}_0$, let ${\bf z}^{X}_\tau({\bf z}_0)$ denote the final conditions, after implementation of the cyclic protocol corresponding to measurement outcome $X \in \{A,B,C,D\}$. 
The work performed on the system as it evolves from ${\bf z}_0$ to ${\bf z}^X_\tau({\bf z}_0)$ is given by 
\begin{equation}
\label{eq:worktraj}
W=H_0({\bf z}^X_\tau({\bf z}_0))-H_0({\bf z}_0)
\end{equation}
Over many repetitions of the process, with the protocol $X$ determined by the measurement of initial energy, the average work performed on the system is 
\begin{equation}
\label{eq:avgwork}
\langle W \rangle = \sum_X^{A,B,C,D} \int_{{\bf z}_0\in X}\, {\rm d}{\bf z}_0 \, p^{\rm eq}({\bf z}_0) \, \left( H_0({\bf z}_\tau^X({\bf z}_0))-H_0({\bf z}_0) \right)
\end{equation}
where $p^{\rm eq} \propto \exp(-\beta H_0)$, and $\int_{{\bf z}_0 \in X}$ indicates integration over all microstates ${\bf z}_0$ that result in the measurement outcome $X$.
Eq.~\ref{eq:avgwork} can be rewritten as 
\begin{equation}
\label{eq:avgwork2}
\langle W \rangle = \beta^{-1} \sum_X^{} \int_{{\bf z}\in X}\, {\rm d}{\bf z} \, p^{\rm eq}({\bf z})\ln\frac{p^{\rm eq}({\bf z})}{p^{\rm eq}({\bf z}^X_\tau({\bf z}))}
\end{equation}
(dropping the subscript $0$).
Let us now define two functions
\begin{align}
\label{condprob}
 f_X({\bf z})  &\equiv
\begin{cases}
p^{\rm eq}({\bf z})/P_X \quad & \text{if ${\bf z} \in X$} \\
0 \quad & \text{if ${\bf z} \notin X$}
\end{cases} \\
\label{eq:gX}
g_X({\bf z}) &\equiv p^{\rm eq}({\bf z}^X_\tau({\bf z}))
\end{align}
where $P_X \equiv \int_{{\bf z} \in X} p^{\rm eq}({\bf z})$ is the probability that the outcome of the measurement is $X$.
We can interpret $f_X({\bf z})$ as the probability distribution of initial microstates, {\it conditioned} on the outcome $X$.
Moreover, $\int {\rm d}{\bf z} \, g_X({\bf z})= \int {\rm d}{\bf z}^X_\tau \, p^{\rm eq}({\bf z}^X_\tau )=1$ (since phase volume is preserved, ${\rm d}{\bf z}= {\rm d}{\bf z}^X_\tau({\bf z})$, by Liouville's theorem), therefore $g_X({\bf z})$ can also be interpreted as a probability distribution on phase space.

With these definitions, Eq.~\ref{eq:avgwork2} becomes
\begin{align}
\langle W \rangle &= \beta^{-1}\sum_X^{} \int \, {\rm d}{\bf z} \, P_X f_X({\bf z}) \ln\frac{P_X f_X({\bf z}) }{g_X({\bf z})}\\
\label{eq:avgwork2_cond}
&= \beta^{-1}\sum_X^{} P_X  \int \, {\rm d}{\bf z} \, f_X({\bf z}) \ln\frac{ f_X({\bf z}) }{g_X({\bf z})} +\beta^{-1}\sum_X^{} P_X \ln P_X 
\end{align}
The integral appearing in Eq.~\ref{eq:avgwork2_cond} is the relative entropy or Kullback-Leibler divergence between the distributions $f_X({\bf z})$ and $g_X({\bf z})$; this quantity is equal to zero if the two distributions are identical and is positive otherwise~\cite{Cover2006}:
\begin{equation}
\int f_X \ln \frac{f_X}{g_X} = D[f_X || g_X] \ge 0
\end{equation}
Thus the first sum on the right side of Eq.~\ref{eq:avgwork2_cond} is non-negative, hence
 \begin{equation}
 \label{eq:boundW}
 \langle W \rangle \geq \beta^{-1} \sum_X P_X \ln P_X = -\beta^{-1} {\cal H}
 \end{equation}
which is equivalent to Eq.~\ref{eq:inequality}, the bound we set out to establish.~\footnote{
In fact, as long as our measurement apparatus has more than one possible outcome $X$, this result will be a strict inequality, since $f_X({\bf z}) = 0 \ne g_X({\bf z})$ for any ${\bf z} \notin X$, hence $D[f_X || g_X] > 0$.}

The above derivation hinges on the non-negativity of relative entropy.
A similar approach has recently been taken to obtain inequalities related to the second law of thermodynamics~\cite{Esposito2010,Hasegawa2010,Takara2010,Esposito2011arXiv}, in situations when the system of interest does not necessarily begin (or end) in states of thermal equilibrium.
(See also Ref.~\cite{Jarzynski1999a} for an alternative derivation of such inequalities.)

While the calculation presented here assumes a measurement apparatus with four possible outcomes, it should be clear that the analysis generalizes to any finite number of energy intervals.
In fact, we can even drop the assumption that the measurement is strictly correlated with energy.
That is, suppose phase space is divided into $N$ regions (not necessarily corresponding to energy intervals) and suppose that when the system is in microstate ${\bf z}$, the measurement apparatus returns a value $X$ that identifies the region of phase space to which that microstate belongs.
Finally, a cyclic protocol is assigned to each possible outcome.
It can be verified by the reader that the steps leading to Eq.~\ref{eq:boundW} (equivalently Eq.~\ref{eq:inequality}) remain valid.

Moreover, to this point we have considered a measurement apparatus that is error-free: if the initial microstate ${\bf z}_0$ belongs in region $X$, then the measurement outcome is necessarily $X$.
Let us now consider a more general situation in which $P(X \vert {\bf z}_0)$ represents the probability that the apparatus outputs the value $X$, when a measurement is performed on a system in microstate ${\bf z}_0$.
In the Appendix we analyze this scenario and derive the bound
\begin{equation}
\label{MIbound}
\langle W_{\rm extracted} \rangle \le 
\beta^{-1} {\cal I} 
\end{equation}
where
${\cal I}$
is the {\it mutual information}~\cite{Cover2006} between the variable ${\bf z}_0$ and $X$. 
For error-free measurements (e.g. Eq.~\ref{eq:measure_K=4}), ${\cal I} = {\cal H}$ and Eq.~\ref{MIbound} reduces to Eq.~\ref{eq:boundW}.
When the apparatus is capable of making errors, then ${\cal I} < {\cal H}$~\cite{Cover2006}, which conforms nicely to the intuition that an error-prone measuring device degrades our ability to extract work from the system.
In either case Eq.~\ref{eq:inequality} remains valid.

Finally, we note that the results derived in this section can be generalized to systems evolving according to stochastic equations of motion~\cite{Vaikuntanathan2011}.

\section{Discussion and Conclusions}
\label{sec:discussion}

The past few years have seen considerable interest in the thermodynamics of small systems and in the applicability of the second law to various nanoscale scenarios (see Ref.~\cite{Jarzynski2011} for a recent review), including those involving feedback.
Motivated by the recent work of Marathe and Parrondo~\cite{Marathe2010}, we have studied a model single-particle system that is ``cooled'' under the quasi-static cycling of external parameters, when initial conditions are sampled microcanonically.
We have used this model to construct a procedure for systematically harvesting energy from a thermal reservoir and converting that energy to work, in seeming violation of the Kelvin-Planck statement of the second law.
This procedure, however, involves the repeated measurement of the energy of the system.
Modeling the measurement apparatus in Sec.~\ref{sec:erasure}, we have shown by explicit calculation that the average work delivered per operating cycle does not exceed the average work that must eventually be expended (in accordance with Landauer's principle) to erase the information acquired in the act of measuring the initial energy.
Thus on balance the Kelvin-Planck statement remains satisifed.

Our model illustrates the idea -- which traces back to Maxwell and Szilard -- that knowledge about the microscopic state of a system can be exploited to circumvent the second law of thermodynamics, loosely speaking~\cite{Leff2003}.
In this setting, Eq.~\ref{eq:boundW} places a bound on the work that can be extracted during a cyclic process, following a measurement that provides information about the initial state of the system.
As already mentioned, similar bounds have been obtained and studied in the past few years, both for quantum systems~\cite{Zurek2003,Sagawa2008,Jacobs2009,Kim2011} and for systems evolving according to stochastic equations of motion~\cite{Touchette2000,Kim2007,Cao2009,Suzuki2009,Sagawa2010,Ponmurugan2010,Fujitani2010,Horowitz2010,Toyabe2010,Abreu2011}.
We also note that Eq.~\ref{eq:avgwork2_cond}, a precursor to Eq.~\ref{eq:boundW}, generalizes the relative entropy work relation of Kawai, Parrondo and Van den Broeck~\cite{Kawai2007} to processes with feedback.

Let us now return to a point mentioned in Sec.~\ref{sec:model}: the apparent incompatibility of Eq.~\ref{eq:Ef} with Liouville's theorem.
Consider a single {\it energy shell}, that is the set of all points ${\bf z}_0$ with a particular value of energy $E_i = H_0({\bf z}_0)$.
This set, which we denote ${\cal S}_i$, has the topology of a simple closed loop in phase space.
Let us assume that this energy shell is located in region $II$, hence $E_1<E_i<E_2$.
If we evolve trajectories from initial conditions in ${\cal S}_i$, using the protocol in Fig.~\ref{fig:symmetricProtocol}, we arrive at a set of final conditions, ${\cal S}_f$, which also has the topology of a simple closed loop: 
\begin{equation}
{\cal S}_i = \{ {\bf z}_0 \, \vert \, H_0({\bf z}_0) = E_i \}
\quad \rightarrow \quad
{\cal S}_f = \{ {\bf z}_\tau({\bf z}_0) \, \vert \, H_0({\bf z}_0) = E_i \}
\end{equation}
By Liouville's theorem, these loops enclose equal volumes of phase space:
$\Omega[{\cal S}_f] = \Omega[{\cal S}_i]$.
This, however, is incompatible with a literal interpretation of Eq.~\ref{eq:Ef}, which seems to assert that {\it every} initial condition with energy $E_i$ leads to a net decrease of energy, $E_f < E_i$, in other words that ${\cal S}_f$ is contained entirely in the interior of ${\cal S}_i$.
To address this apparent contradiction, we sketch a more careful interpretation of Eq.~\ref{eq:Ef}.

For any finite duration $\tau$, there exist {\it some} initial conditions ${\bf z}_0 \in {\cal S}_i$ that yield trajectories for which the system's energy increases: $H_0({\bf z}_\tau({\bf z}_0)) > E_i$.
We will refer to these trajectories as ``bad actors'', as they spoil the picture shown in Fig.~\ref{fig:sequence}.~\footnote{
In simulations, we have observed bad actors that begin near the bottom of region $II$, but get trapped in the right well at the end of the first stage of the process, e.g.\ just before $t=\tau/4$ in Fig.~\ref{fig:sequence}.
As a result, they do not get drawn into the left well during the second stage.
They subsequently ``float'' on top of the darkly shaded set $I$ in Fig.~\ref{fig:sequence}, and end the process with $H_0({\bf z}_\tau({\bf z}_0)) \approx E_2$.
}
While bad actors exist for any finite $\tau$, the probability to generate one of these trajectories generally decreases with increasing $\tau$, for initial conditions sampled microcanonically from ${\cal S}_i$.
We have observed this trend in numerical simulations over a range from $\tau=1200$ to 2000 (data not shown);
and as mentioned in Sec.~\ref{sec:model}, for $E_i=2.8$ and $\tau=12000$ no bad actors were observed among $10^5$ trajectories.
Thus for large but finite $\tau$, we expect ${\cal S}_f$ to be a highly convoluted, closed loop -- necessarily enclosing the same volume of phase space as ${\cal S}_i$ -- with much of the loop concentrated at low energies near the value predicted by Eq.~\ref{eq:Ef}, but with tendrils reaching into the region of energies higher than than $E_i$. 
We believe this issue deserves a more careful treatment, but this is beyond the scope of the present paper.
We end with a conjecture regarding the quasi-static limit:
\begin{equation}
\lim_{\tau\rightarrow\infty} P \left[ \left\vert H_0({\bf z}_\tau) - E_f \right\vert < \frac{\epsilon}{2} \right] = 1
\quad \text{for any $\epsilon > 0$}
\end{equation}
where the quantity inside the limit is the probability to generate a trajectory whose final energy falls within an interval of width $\epsilon$ around the value predicted by Eq.~\ref{eq:Ef}, and microcanonical sampling at energy $E_i$ is assumed.
We believe this conjecture represents the proper way to understand the validity of Eq.~\ref{eq:Ef} and Fig.~\ref{fig:sequence}.
Similar comments apply to Eq.~\ref{eq:Ef_asymm} and Fig.~\ref{fig:asymmetricSequence}.

Our results suggest several avenues for future research.

First, it would be interesting to explore a quantum-mechanical version of our model system.
Here, the possibility of tunneling between the left and right wells introduces a new aspect to the problem, possibly spoiling the picture developed in Sec.~\ref{sec:model} by preventing particles from getting trapped.

Because the protocols discussed in Sec.~\ref{sec:model} involve the quasi-static cycling of external parameters, it is natural to wonder whether the swapping of regions $I$ and $II$ (illustrated in Fig.~\ref{fig:sequence}) can be described in terms of a geometric phase.

Finally, we have not explicitly modeled the ``demon'' in Sec.~\ref{sec:erasure}.
Instead, we have assumed the existence of some mechanism by which a particular outcome of the measurement leads to the implementation of the corresponding protocol.
It would be interesting, however, to model this mechanism explicitly within a Hamiltonian framework, either by introducing additional degrees of freedom to model the demon or by specifying coupling terms between the measurement device and the system.
In this case, we anticipate that the bound on extracted work will be given in terms of the correlation between the state of the system and the state of the measuring device and/or demon~\cite{Touchette2000,Zurek2003,Jacobs2009,Cao2009}.

\acknowledgements
We gratefully acknowledge useful discussions and correspondence with Eric Heller, Jordan Horowitz, Daniel Lathrop, Rahul Marathe, Juan Parrondo and Wojciech Zurek, as well as financial support from the National Science Foundation (USA) under grants CHE-0841557 and DMR-0906601, and the University of Maryland, College Park.

\appendix
\section{Analysis of error-prone measurement devices}
\label{Appendix:1}

Consider a measurement apparatus with a discrete set of possible outputs, $X = A, B, C, \cdots$, and let $P(X \vert {\bf z}_0)$ denote the probability to obtain outcome $X$, when the measurement is performed on a system in microstate ${\bf z}_0$.
We assume that every measurement produces some outcome, hence $\sum_X P(X \vert {\bf z}_0) = 1$ for any ${\bf z}_0$.
As before, a cyclic protocol is chosen based on the outcome of the measurement.
For initial conditions ${\bf z}_0$, let ${\bf z}^{X}_\tau({\bf z}_0)$ denote the final conditions, after implementation of the protocol corresponding to outcome $X$.
The work performed on the system is given by Eq.~\ref{eq:worktraj}, and averaging over many repetitions of the process gives us 
\begin{align}
\label{eq:avgwork_A}
\langle W \rangle &= \int {\rm d}{\bf z}_0 \, p^{\rm eq}({\bf z}_0) \sum_X \, P(X \vert {\bf z}_0) \, \left( H_0({\bf z}_\tau^X({\bf z}_0))-H_0({\bf z}_0) \right) \\
&= \beta^{-1}
\sum_X^{} \int {\rm d}{\bf z}_0 \, P({\bf z}_0,X) \, 
\ln\frac{p^{\rm eq}({\bf z}_0)}{p^{\rm eq}({\bf z}^X_\tau({\bf z}_0))}
\end{align}
where $P({\bf z}_0,X)$ is the joint probability that the system is initially in microstate ${\bf z}_0$ and the measurement outcome is $X$.
Dropping the subscript $0$, we now introduce two probability distributions (compare with Eqs.~\ref{condprob}, \ref{eq:gX})
\begin{align}
 f_X({\bf z})  &\equiv P({\bf z} \vert X) = P({\bf z},X) / P_X \\
g_X({\bf z}) &\equiv p^{\rm eq}({\bf z}^X_\tau({\bf z}))
\end{align}
where $P_X = \int {\rm d}{\bf z} \, P({\bf z},X)$ is the net probability to generate the outcome $X$, and $P({\bf z}|X)$ denotes the conditional probability distribution that the initial microstate is ${\bf z}$, given the measurement outcome $X$.
In terms of these distributions we now have
\begin{align}
\langle W \rangle &= 
\beta^{-1}
\sum_X^{} \int {\rm d}{\bf z} \, P({\bf z},X) \,
\ln \left[
\frac{f_X({\bf z})}{g_X({\bf z})} \cdot \frac{P_X \, p^{\rm eq}({\bf z})}{P_X \, f_X({\bf z})}
\right] \\
\label{eq:avgwork3_A}
&= 
\beta^{-1} \sum_X^{} P_X \int {\rm d}{\bf z} \, f_X({\bf z}) \ln\frac{f_X({\bf z})}{g_X({\bf z})} -\beta^{-1}  \sum_X^{} \int {\rm d}{\bf z} \, P({\bf z},X) \ln\frac{P({\bf z},X)}{p^{\rm eq}({\bf z}) P_X}
\end{align}
On the last line, the first term is a relative entropy, and therefore non-negative;
while the second term (apart from the factor $\beta^{-1}$) is the mutual information between ${\bf z}$ and $X$.
We thus arrive at
\begin{equation}
\label{eq:MIbound_A}
\langle W \rangle \ge -\beta^{-1} {\cal I}
\end{equation}
equivalently Eq.~\ref{MIbound}.

%

\end{document}